\documentclass[12pt,preprint]{aastex}

\shorttitle{Intrinsic polarization angle ambiguity in Faraday tomography}
\shortauthors{Kumazaki, Akahori, Ideguchi, Kurayama, and Takahashi}

\begin{document}
\title{Properties of Intrinsic Polarization Angle Ambiguities in Faraday Tomography}
\author{Kohei Kumazaki$^1$, Takuya Akahori$^2$, Shinsuke Ideguchi$^3$, Tomoharu Kurayama$^4$, and Keitaro Takahashi$^3$}
\affil{$^1$Nagoya University, Furo-cho, Chikusa-ku, Nagoya 464-8601, Japan;kumazaki@a.phys.nagoya-u.ac.jp\\
$^2$Sydney Institute for Astronomy, School of Physics, The University of Sydney, NSW 2006, Australia: akahori@physics.usyd.edu.au\\
$^3$University of Kumamoto, 2-39-1, Kurokami, Kumamoto 860-8555, Japan; 121d9001@st.kumamoto-u.ac.jp, keitaro@sci.kumamoto-u.ac.jp\\
$^4$Center for Fundamental Education, Teikyo University of Science, 2525 Yatsusawa, Uenohara, Yamanashi 409-0193, Japan; kurayama@ntu.ac.jp}

\begin{abstract}

Faraday tomography is a powerful method to diagnose polarizations and
Faraday rotations along the line of sight. The quality of Faraday
tomography is, however, limited by several conditions. Recently, it is
reported that Faraday tomography indicates false signals in some
specific situations. In this paper, we systematically investigate the
condition of the appearance of false signals in Faraday tomography. We
study this by pseudo-observing two sources within a beam, and
change in the intrinsic polarization angles, rotation measures,
intensities, and frequency coverage. We find that false signals arise
when rotation measure between the sources is less than 1.5 times the
full width at half maximum of the rotation measure spread
function. False signals also depend on the intensity ratio between the
sources and are reduced for large ratio. On the other hand, the
appearance of false signals does not depend on frequency coverage,
meaning that the uncertainty should be correctly understood and taken
into consideration even with future wide-band observations such as Square
Kilometer Array (SKA).

\end{abstract}


\section{Introduction}
Cosmic magnetic fields are interesting topics for modern astrophysics
and cosmology, because magnetic fields could affect the formation of
cosmic structures and imprint themselves on histories and properties of
cosmic evolutions. For example, magnetic fields are essential for
instabilities in galactic gaseous disks \citep[e.g.,][]{mac13}, play
crucial roles in structures, particle accelerations, and radio
emissions in galaxy clusters \citep[e.g.,][]{tak08,fuj13}, record
evolutions of turbulence in the cosmic web \citep[e.g.,][]{ar10,ar11},
and affect cosmic ray or $\gamma$-ray propagation such as pair echo
\citep{tkm09,ner10,tak12,tak13}.

Faraday rotation, the rotation of the polarization angle of
electromagnetic waves traveling in magnetized plasma, is a standard
tool to measure Galactic and extragalactic magnetic fields
\citep{sch09,wol10,nou12,bre11,gae05,bec09,fle11}. For a line of sight
(LOS) toward a single polarized source, Faraday rotation can be
written as
\begin{equation}\label{eq1}
\chi(\lambda^2) = \chi_0 + \phi \lambda^2, 
\end{equation}
where $\lambda$ is the wavelength, $\chi$ and $\chi_0$ are the
rotation angles at the observer and the source, respectively, and
$\phi$ is the Faraday depth or Faraday rotation measure (RM) which is
an integration of the LOS component of magnetic fields from the source
to the observer with a weight of electron density.
Notice that "Faraday depth" does not correlate with physical distance
of source and sign which shows the direction of magnetic field toward
observer dramatically changes the behavior of polarization angle
and Faraday dispersion function (see Section 2).

Faraday depth can be estimated based on the linear relation between
$\lambda^2$ and $\chi$ in equation (\ref{eq1}). The estimation,
however, becomes uncertain if there exists multiple sources within a
beam, in the case $\lambda^2$ and $\chi$ does not display a linear
relation \citep[see e.g.,][]{bd05}. This problem is generally
inevitable in studies of extragalactic magnetic fields, since we
observe extragalactic sources viewed through Galactic emissions
\citep{bre11}. Therefore, we need more sophisticated methods to
diagnose polarized emissions and Faraday depths along the LOS.

One of the sophisticated methods is so-called RM synthesis or Faraday
tomography proposed in radio astronomy
\citep{bu66,bd05,and12,bec12,bel12,and13}, which has been applied by
many authors \citep{hea09,sch09,wol10,fri11,iac13}. This method
transforms polarization spectra as a function of wavelength into those
as a function of Faraday depth. For example, \cite{osu12} analyzed
point sources with Faraday tomography and spectral fitting with
Faraday rotation models. They found that some of their polarization
spectra could not be fitted with a single-source model but are well
fitted with a multiple-source model. This means that they resolved the
point sources by their polarization spectra. This demonstrates that
Faraday tomography is effective for cases with multiple sources within
a single beam.

Since the quality of Faraday tomography primarily depends on frequency
coverage of the data, Faraday tomography is expected to become a major
tool of radio polarimetry using future facilities such as Square
Kilometer Array (SKA). Recently, however, \cite{far11} reported some
ambiguities associated with Faraday tomography. They considered two
sources located very closely each other in Faraday depth, and varied
the difference of intrinsic polarization angles of the two
sources. They found that Faraday tomography could resolve two
components correctly in some cases but, in other cases, it erroneously
detected three components including false signals. This phenomenon,
called RM ambiguity, would substantially degrade Faraday tomography
unless we understand it.

In this paper, we extend the previous work and examine the conditions
of the appearance of false signals more systematically. We investigate
the dependence of false signals on intrinsic polarization angles,
rotation measures, and intensities of the two sources. We also change
frequency coverage based on planned observations with Australian SKA
Pathfinder (ASKAP) and SKA. In section 2, we describe our model and
calculations. We study the conditions of false signals in section 3,
and give a summary and discussion in section 4.

\section{Model and Calculation}
We follow a basic formula of Faraday tomography described by
\cite{bd05} and summarized by \cite{aktr12} and \cite{ide13}. We start
with the polarized intensity, as an observable quantity, which is a
complex function in the square of the wavelength:
\begin{eqnarray}\label{eq2}
P(\lambda^2)
&=& pI(\lambda^2)e^{2i\chi(\lambda^2)} \nonumber\\
&=& Q(\lambda^2) + iU(\lambda^2) \nonumber\\
&=& \int_{-\infty}^{\infty} F(\phi)e^{2i\phi\lambda^2}d\phi, 
\end{eqnarray}
where $p$ is the polarization degree, $I$, $Q$ and $U$ are the Stokes
parameters. $F(\phi)$ is the Faraday dispersion function (FDF), which
is a complex polarized intensity as a function of Faraday depth. The
last equation can be inverted as
\begin{equation}\label{eq3}
F(\phi) = \int_{-\infty}^{\infty} P(\lambda^2) e^{-2i\phi\lambda^2} d\lambda^2.
\label{eq:FDF}
\end{equation}

The FDF is hence obtained from the polarized intensity, though its
reconstruction is not perfect because negative $\lambda^2$ is
unphysical and observable wavelengths are limited. We introduce the
window function $W(\lambda^2)$ which has nonzero value at observed
$\lambda^2$ and zero elsewhere including negative $\lambda^2$. The
observed complex polarized intensity and the reconstructed FDF can be
written as
\begin{equation}\label{eq4}
\tilde P(\lambda^2) = W(\lambda^2) P(\lambda^2),
\end{equation}
and
\begin{equation}\label{eq5}
\tilde F(\phi) = R(\phi)\ast F(\phi), 
\end{equation}
respectively, where we used the convolution theorem and $\ast$ indicates convolution.
\begin{equation}\label{eq6}
R(\phi)
= \int^{\infty}_{-\infty} W(\lambda^2)e^{-2i\phi\lambda^2}d\lambda^2
\end{equation}
is the rotation measure spread function (RMSF), which is defined with
the inverse Fourier transform of $W(\lambda^2)$.

\begin{figure}[tp]
\figurenum{1}
\epsscale{1.0}
\includegraphics[width=160mm]{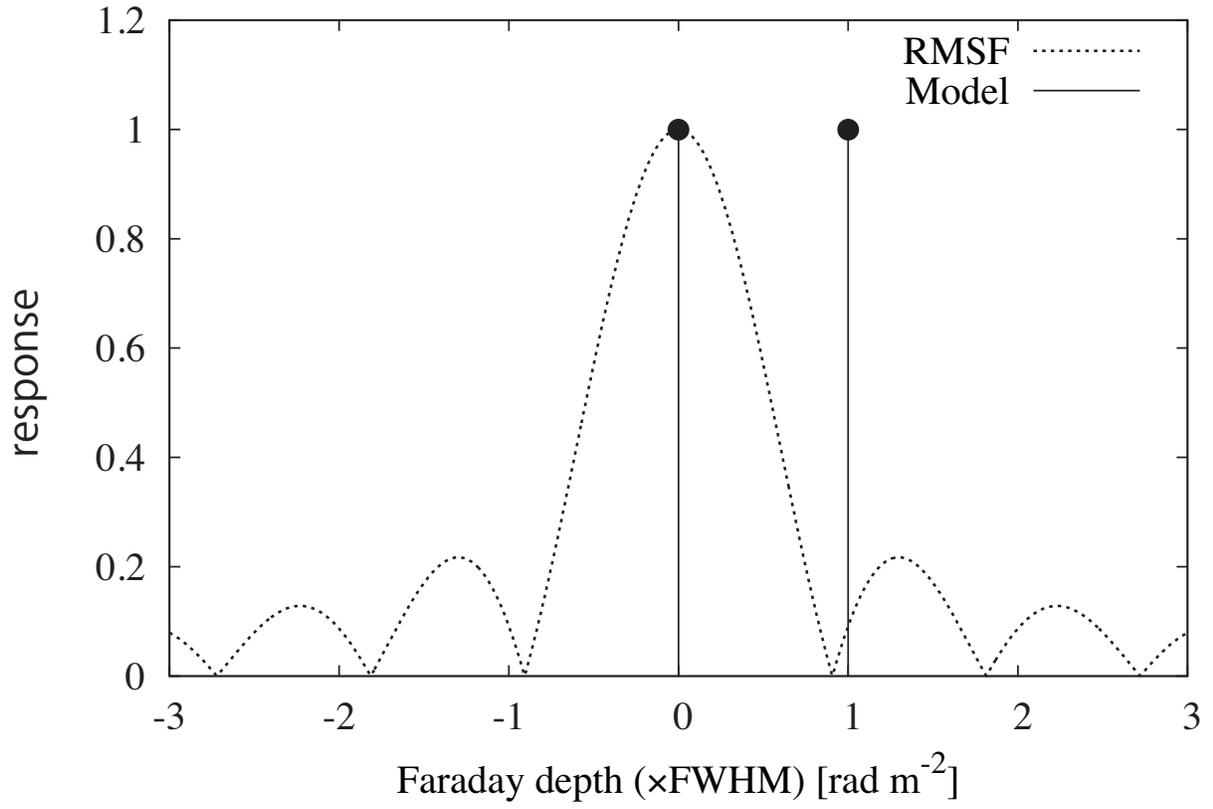}
\caption{ The rotation measure spread function (RMSF) and an example
  of the model of Faraday dispersion function which is composed of 2
  point sources. \label{f1} }
\end{figure}

Equation (\ref{eq5}) indicates that the quality of reconstruction of
$F(\phi)$ is determined by $R(\phi)$. Figure \ref{f1} shows the RMSFs,
where the full width at half maximum (FWHM) is determined by the
observed wavelength as
\begin{equation}\label{eq7}
{\rm FWHM}=\frac{2\sqrt{3}}{\Delta\lambda^2}
\end{equation}
for a single continuous and uniform weighting band,
$\Delta\lambda^2\equiv \lambda^2_{\rm max} - \lambda^2_{\rm min}$,
where $\lambda_{\rm max}$ and $\lambda_{\rm min}$ are the largest and
shortest wavelengths of the observations, respectively. Hereafter, we
test the cases for ASKAP and SKA. The FWHM is 22.26 rad/m$^2$ for
ASKAP and 0.189 rad m$^{-2}$ for SKA, based on the proposed frequency
coverage listed in Table \ref{tab1}.

\placefigure{f1}

%%% table1 here %%%

Due to finite width of frequency coverage, we see in Figure \ref{f1}
that the RMSF has sidelobes, and thus the reconstructed FDF obtained
from equation (\ref{eq5}) has artificial fringes due to the
sidelobes. To remove them, we employ RM CLEAN
\citep[e.g.,][]{hea09,bel12}. RM CLEAN is an algorithm similar to the
CLEAN deconvolution developed for reconstruction of images obtained
using a aperture synthesis radio telescope \citep[e.g.,][]{hog74}. We
perform RM CLEAN as follows. First, we seek the peak value of the
reconstructed FDF at $\phi_s$, $F(\phi_s)$, and add a new CLEAN
component with its peak value multiplied by $\gamma$ into a CLEAN
component list. Here $\gamma$ is the gain factor and we adopt
$\gamma=0.1$. Second, we subtract the shifted-scaled RMSF, $\gamma
F(\phi)R(\phi-\phi_s)$, from the reconstructed FDF. Third, we repeat
the above two steps until the $F(\phi_s)$ becomes below 1/1000 of the
peak intensity of the model or until number of iterations is beyond
3000. Forth, we accumulate CLEAN components and multiply the
accumulated CLEAN components by a CLEAN beam. Here, the CLEAN beam is
a Gaussian beam with the FWHM of the RMSF. The multiplied FDF is
called the cleaned FDF. Finally, we add the residual of the
reconstructed FDF to the cleaned FDF.

The source model we adopted is the one studied by Farnsworth et
al. (2011).  In the model, we consider two polarized sources within a
single beam. The two sources are approximated as Faraday thin,
emitters inside which RMs are negligible, and are described as delta
functions in $F(\phi)$. We do not specify real situations, but the model
is rather general in the sense that two Faraday thin sources are
located at certain Faraday depths such as radio lobes/cores
(Farnsworth et al. 2011).  The complex polarized intensity can
be written as
\begin{equation}\label{eq8}
P(\lambda^2_j) = \sum_{k=1}^2 A_k e^{2i(\chi_{0,k} + \phi_k \lambda^2_j)},
\end{equation}
where $A_k$ is the polarized intensity, $\chi_{0,k}$ the intrinsic
polarization angle, and $\phi_k$ is the Faraday depth of the $k$-th
source. We fix $A_1=10.0$, $\phi_1=0.0$ and $\chi_{0,2}=0.0$ as
reference values, and change $A_2$, $\phi_2$, and $\chi_{0,1}$ to
systematically investigate properties of false signals through a
number of case studies. The difference of the intrinsic polarization
angles,
\begin{equation}\label{eq9}
\Delta\chi_0 \equiv \chi_{0, 1} - \chi_{0, 2}
\end{equation}
is, thus, equal to $\chi_{0, 1}$. Although $\Delta\chi_0$ should
  have a value between $-180^\circ$ and $+180^\circ$, it should be
  symmetric for $0^\circ$.  Therefore, we consider only positive
  $\Delta\chi_0$s.   Also, the difference of Faraday depths,
\begin{equation}\label{eq10}
\Delta\phi \equiv \phi_2 - \phi_1
\end{equation}
is equal to $\phi_2$. Fig.~\ref{f1} shows an example of the model FDF
with $A_2 = 10.0$ and $\phi_2 =$ 1.0 FWHM.
It should be noted that the first source with $\phi_1 = 0$ is not necessarily
located closer to the observer compared with the second one, because
the Faraday depth is not a monotonic function of the physical distance in general.

Hereafter, we do not take into account the noise in order to confirm
that false signals are not coming from the noise effect but inherent
in this method.

\section{Results}
\subsection{Difference of Intrinsic Polarization Angles, $\Delta\chi_0$}

\begin{figure}[tp]
\figurenum{2}
\epsscale{1.0}
\includegraphics[width=160mm]{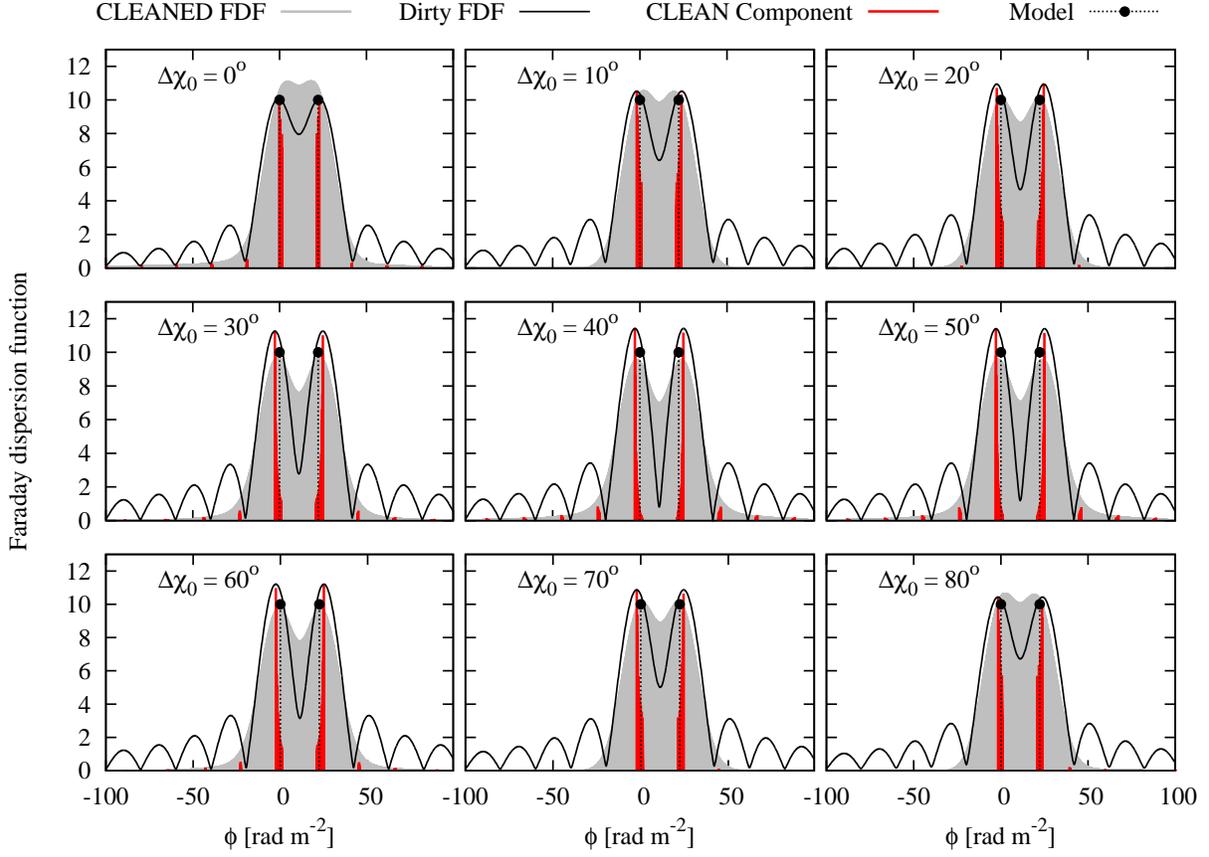}
\caption{ Faraday dispersion function (FDF) for $\Delta\chi_0=$ 0 - 80
  degree, $A_2/A_1 = 1.0$, and $\Delta\phi = 22.26$ rad m$^{-2}$ (1.0
  FWHM) for the case with ASKAP. The black dot-dashed and solid lines
  represent the model and reconstructed FDFs, respectively. The shaded
  area in gray shows the cleaned FDF calculated by RM CLEAN
  algorithm. The red lines are the accumulated CLEAN
  components. \label{f2} }
\end{figure}

\placefigure{f2}

\begin{figure}[tp]
\figurenum{3}
\epsscale{1.0}
\includegraphics[width=160mm]{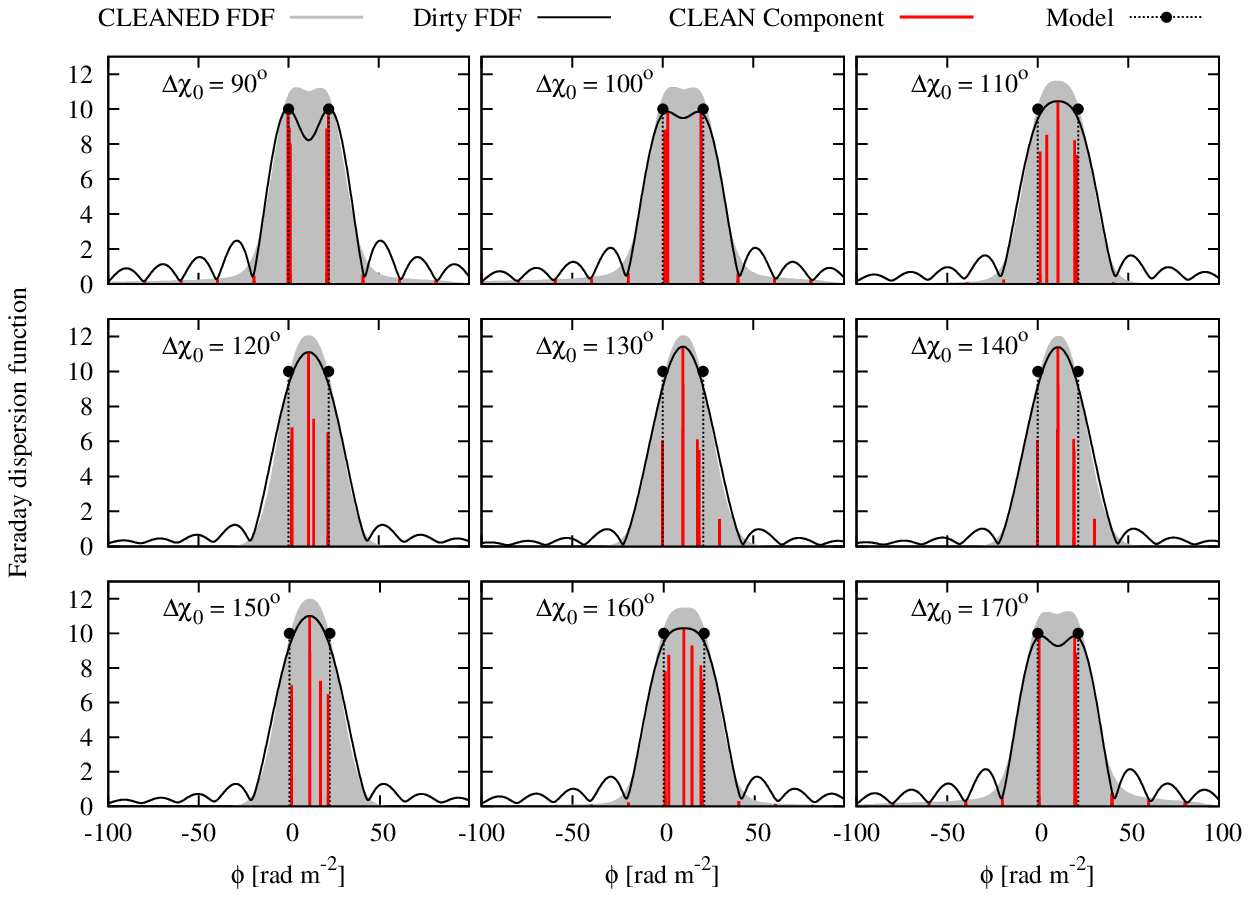}
\caption{
Same as Figure \ref{f2} but for $\Delta\chi_0=$ 90 - 170 degree. \label{f3}
}
\end{figure}

\placefigure{f3}

In figures \ref{f2} and \ref{f3}, we show the results of Faraday
tomography and RM CLEAN for the cases with different $\Delta\chi_0$,
where $A_2/A_1 = 1.0$ and $\Delta\phi = 22.26$ rad m$^{-2}$ (1.0 FWHM)
for the case with ASKAP are fixed. The two components are correctly
detected except for the cases with $110^\circ \le \Delta \chi_0 \le
160^\circ$. In the cases with $110^\circ \le \Delta \chi_0 \le
160^\circ$, the reconstructed FDF has a single peak around the mean
Faraday depth of the two components and the CLEAN components includes
false signal larger than the correct signals. This phenomenon was
reported by \cite{far11} as RM ambiguity.  One may think that the
result should be symmetric with respect to $\Delta \chi_0 = 90^\circ$,
that is, the result should depend only on the absolute value of the
$\Delta \chi_0$.  Actually, this is not the case and the asymmetry comes
from the relative sign of $\Delta \chi_0$ and $\Delta \phi$.

\cite{far11} indicated that two sources with a close separation
$\Delta\phi \lesssim$ 1.0 FWHM would be detected as a single, false
source, which is actually confirmed with Figure \ref{f3}. On the other
hand, there has not been reported that false signals arise for the
cases with $\Delta\phi$ sufficiently larger than the FWHM. In order to
make clear the border of RM ambiguity, in the next subsection we study
the cases with $\Delta\phi \gtrsim$ 1.0 FWHM and do not consider the
cases with $\Delta\phi \ll$ 1.0 FWHM.

\subsection{Separation between components, $\Delta\phi$}

\begin{figure}[tp]
\figurenum{4}
\epsscale{1.0}
\includegraphics[width=160mm]{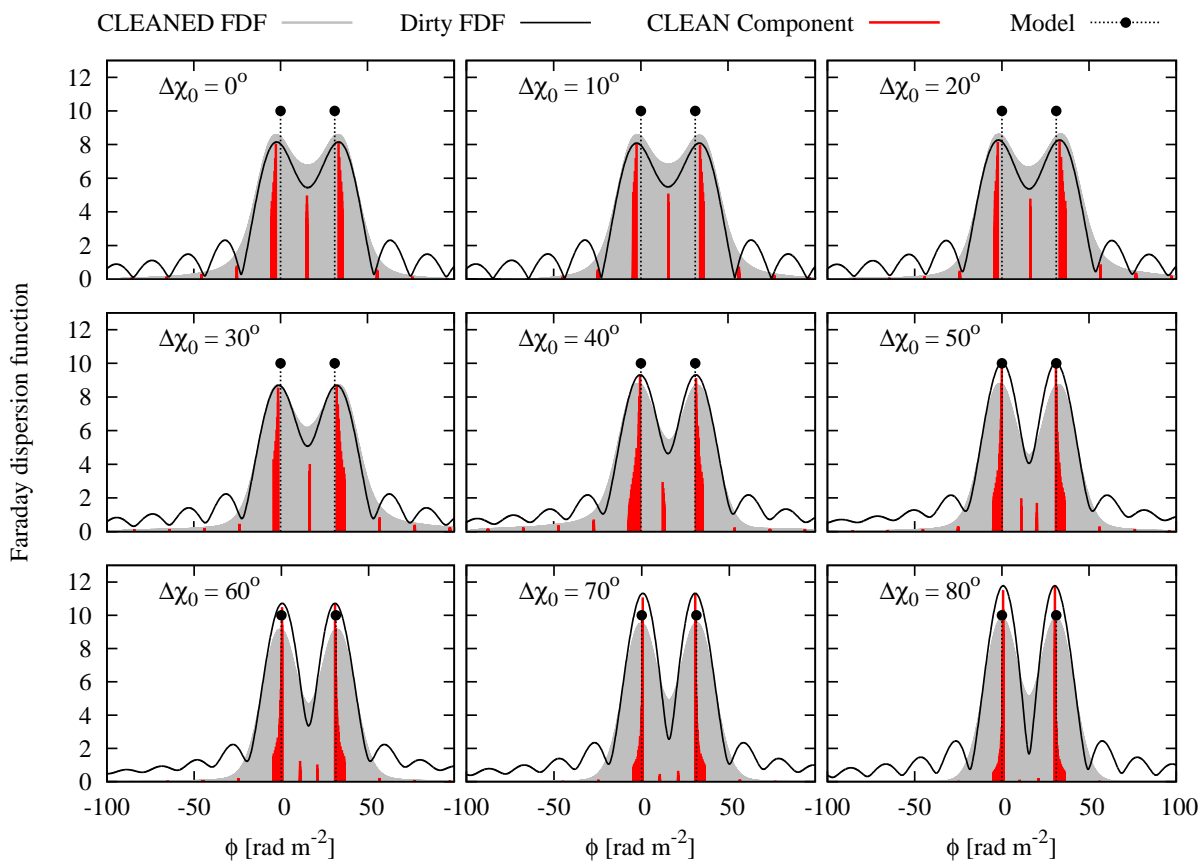}
\caption{ Faraday dispersion function (FDF) for $\Delta\chi_0=$ 0 - 80
  degree, $A_2/A_1 = 1.0$, and $\Delta\phi = 31.16$ rad m$^{-2}$ (1.4
  FWHM) for the case with ASKAP. Descriptions of lines and colors are
  the same as Figure \ref{f2}. \label{f4} }
\end{figure}

\placefigure{f4}

\begin{figure}[tp]
\figurenum{5}
\epsscale{1.0}
\includegraphics[width=160mm]{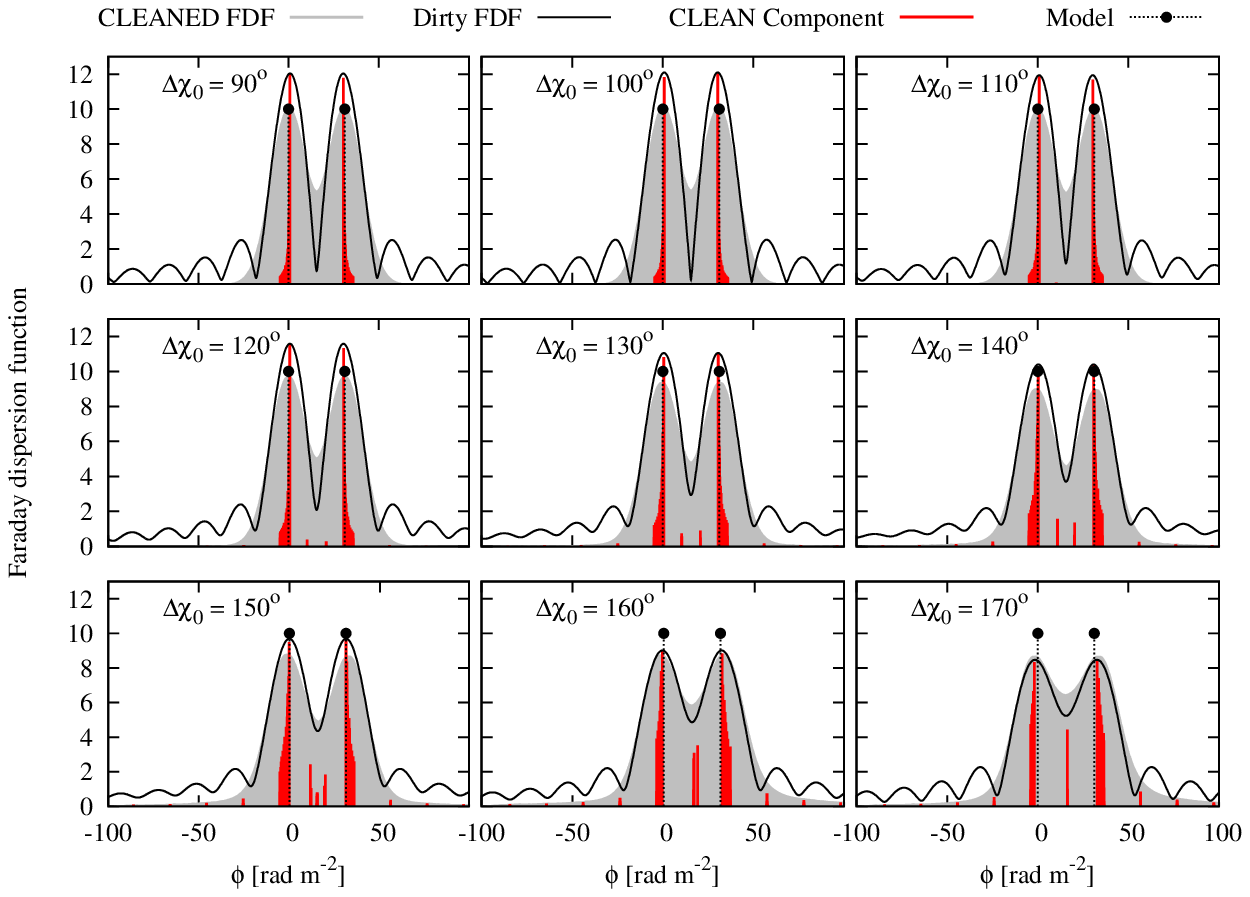}
\caption{
Same as Figure \ref{f4} but for $\Delta\chi_0=$ 90 - 170 degree. \label{f5}
}
\end{figure}

\placefigure{f5}

Figures \ref{f4} and \ref{f5} show the FDFs for the cases with
$\Delta\phi = 1.4$ FWHM. Here, for quantification, we define the false
signals to be CLEAN components arisen in Faraday depth from
$\Delta\phi/4$ to $3\Delta\phi/4$ with an amplitude larger than half
of that of the largest CLEAN component. In all cases, the
reconstructed FDF and cleaned FDF have two peaks near the input
Faraday depth and the CLEAN components located at the correct
positions are dominant. However, false signals can be seen in the
cases with $0 \le \Delta \chi_0 \le 20$ and $\Delta \chi_0 =
170$. Thus, false signals appear even for a source separation larger
than the FWHM of the RMSF.

\begin{figure}[tp]
\figurenum{6}
\epsscale{1.0}
\includegraphics[width=160mm]{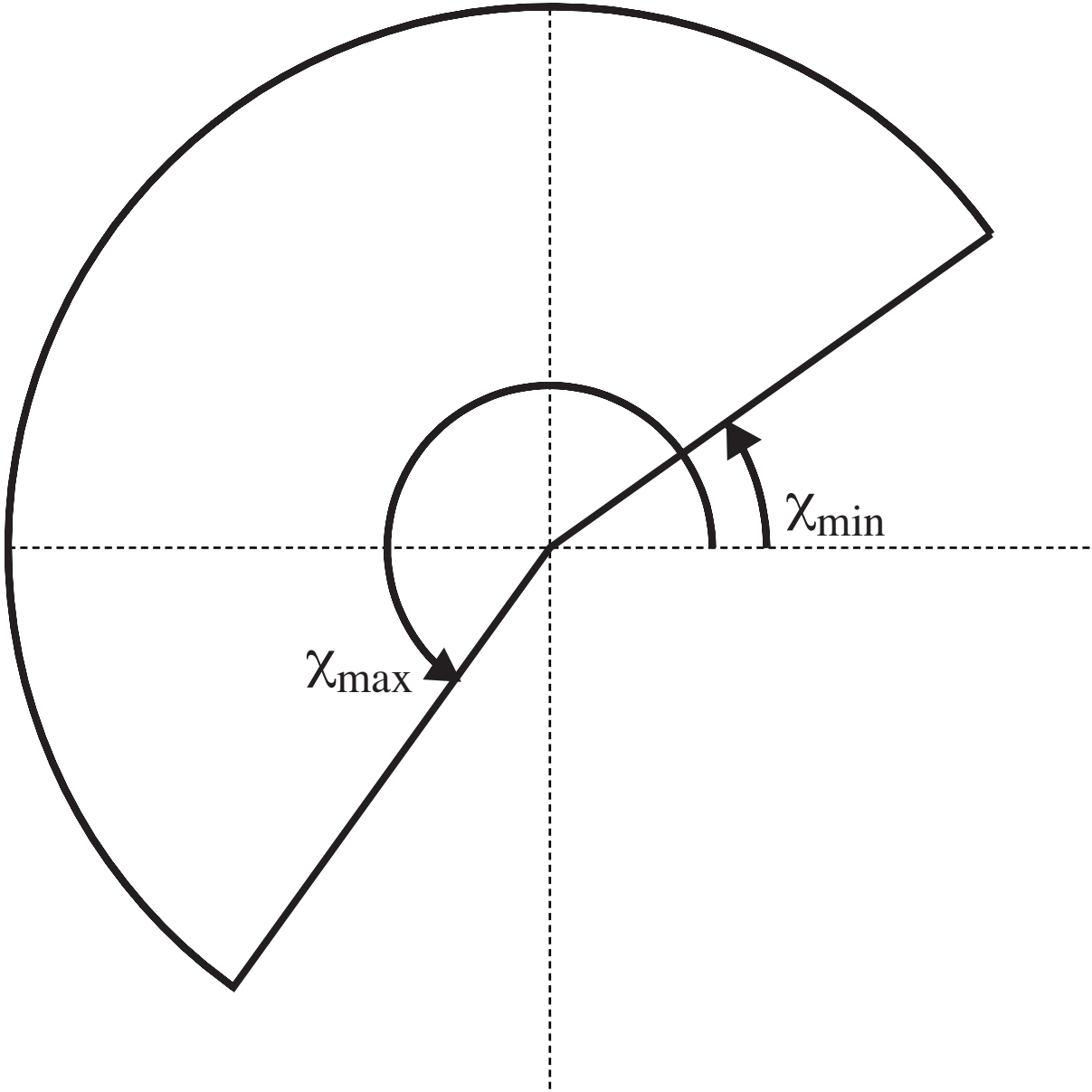}
\caption{ Definition of the pie display for rotation angles of light
  emitted by the second source ($k=2$). $\chi_{\rm max}=\Delta\phi
  \lambda^2_{\rm max}$ and $\chi_{\rm min}=\Delta\phi \lambda^2_{\rm
    min}$ are the rotation angles at the first source used in
    Figure~\ref{f7} and \ref{f10}.\label{f6} }
\end{figure}

\placefigure{f6}

To understand how large separation is needed to avoid false signals,
we investigate false signals for the cases with various values of
$\Delta\phi$, fixing the other parameters.  For systematic displays,
we show the rotation angles of light emitted by the second source
($k=2$) as a sector shaped with a thick line shown in Figure \ref{f6},
where $\chi_{\rm max}=\Delta\phi \lambda^2_{\rm max}$ and $\chi_{\rm
  min}=\Delta\phi \lambda^2_{\rm min}$ are the rotation angles for
$\lambda_{\rm max}$ and $\lambda_{\rm min}$ at the first source.

\begin{figure}[tp]
\figurenum{7}
\epsscale{1.0}
\includegraphics[width=160mm]{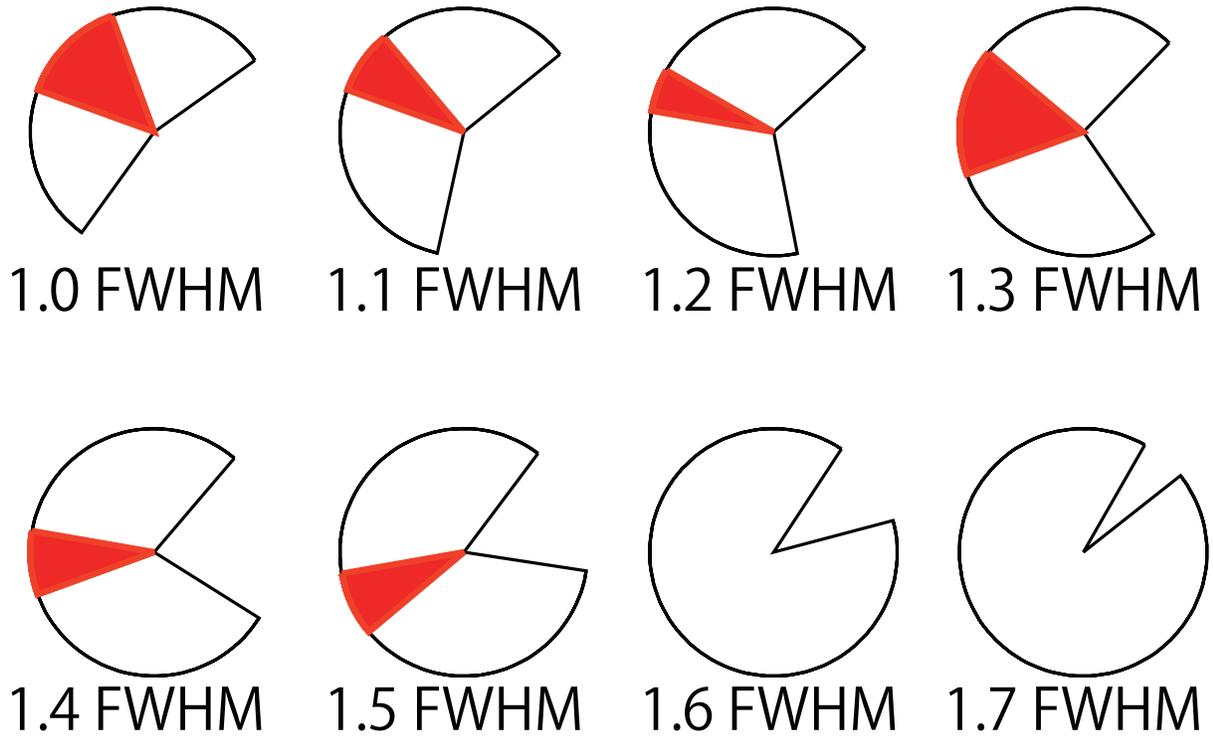}
\caption{ Appearance of false signals. The black sectors are the same
  as Figure \ref{f6} but for different FWHMs. The red sectors
  represent the range of $\Delta\chi_{0}$ where false signals appear
  in the CLEAN components. \label{f7} }
\end{figure}

\placefigure{f7}

Figure \ref{f7} summarize the results for $\Delta\phi = 1.0 - 1.7$
FWHM. The red sectors represent the range of $\Delta\chi_{0}$ where
false signals appear in the CLEAN components. We find that false
signals tend to appear when $\Delta\chi_{0}\sim (\Delta\phi
\lambda^2_{\rm min} + \Delta\phi \lambda^2_{\rm max})/2$. This would
be understandable because the resultant polarizations emitted from the
two sources are similar each other in this case, and thus it is rather
difficult to separate the both sources correctly. We also find that
the range of $\chi_{0, 1}$ which induces false signals does not become
narrow monotonically as $\Delta\phi$ increases.

\begin{figure}[tp]
\figurenum{8}
\epsscale{1.0}
\includegraphics[width=160mm]{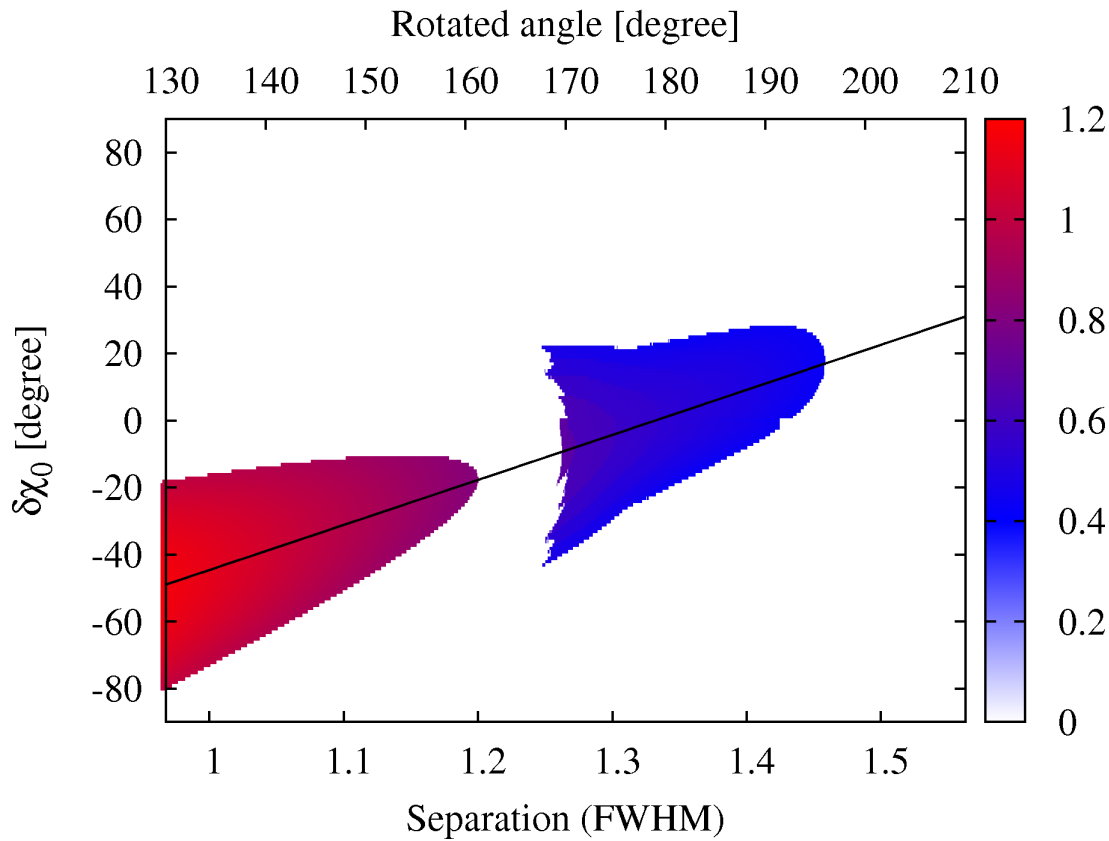}
\caption{ Amplitude of false signals in the CLEAN components. Colors
  show the amplitudes for different separations ($\Delta\phi$) and
  different intrinsic polarization angles ($\delta\chi_0$). Amplitude
  is shown as the ratio between amplitudes of the first and second
  largest CLEAN components, and false signals with the amplitude less
  than 0.5 are not shown. The black solid line represents the rotation
  angle of light at the first source with the wavelength $
  (\lambda^2_{\rm min} + \lambda^2_{\rm max})/2$ emitted by the second
  source ($k=2$). \label{f8} }
\end{figure}

\placefigure{f8}

Figure \ref{f8} shows the amplitude of the false signals for various
$\Delta\phi$ and $\Delta \chi_0$, where we use $\delta \chi_0$,
instead of $\Delta \chi_0$, defined as,
\begin{eqnarray}\label{eq11}
\delta\chi_0 =\left\{
\begin{array}{ll}
  \Delta\chi_0 & {\rm for} \quad \Delta \chi_0 \leq 90^\circ\\
  \Delta\chi_0 -180^\circ &  {\rm for} \quad \Delta \chi_0 > 90^\circ.
\end{array}
\right.
\end{eqnarray}
We can see that the region in which we see false signals is along the
black solid line, which is the track for the light satisfying $
(\lambda^2_{\rm min} + \lambda^2_{\rm max})/2$ emitted by the second
source, i.e. false signals tend to appear when $\Delta\chi_{0}\sim
(\Delta\phi \lambda^2_{\rm min} + \Delta\phi \lambda^2_{\rm max})/2$
as seen in Figure \ref{f7}. We can see that the false signals are
larger than the correct signals for $\Delta\phi < 1.1~{\rm FWHM}$. The
false signals become weaker for larger $\Delta\phi$ but continue to
appear up to $\Delta\phi = 1.45~{\rm FWHM}$, while there is a gap in
$\Delta\phi = 1.2 - 1.25~{\rm FWHM}$. The separation corresponding to
the gap is equal to the location of the second peak of the RMSF
(Figure \ref{f1}) whose amplitude is about 20\% of that of the main
peak. Thus, this gap is considered to be generated by the sidelobe,
which enhances the other source and makes the detections easier, by
accident.

\subsection{Intensity ratio, $A_2/A_1$}

We next investigate the dependence of false signals on the intensity
ratio between the two sources, $A_2/A_1$.  For each $A_2/A_1$, we
validate the appearance of false signals in the case with $0^\circ
\leq \Delta\chi_0 \leq 180^\circ$ and $0.8 \leq \Delta\phi\leq 1.5$
FWHM then pick up the worst case.  The appearance of false signals can
be classified into three types. Type (I) is that false signals appear
for some intrinsic polarization angles and they are larger than the
correct signals. Type (II) is that false signals appear for some
intrinsic polarization angles and they are smaller than the correct
signals. And type (III) is that there is no false signals for any
intrinsic polarization angle.

\begin{figure}[tp]
\figurenum{9}
\epsscale{1.0}
\includegraphics[width=160mm]{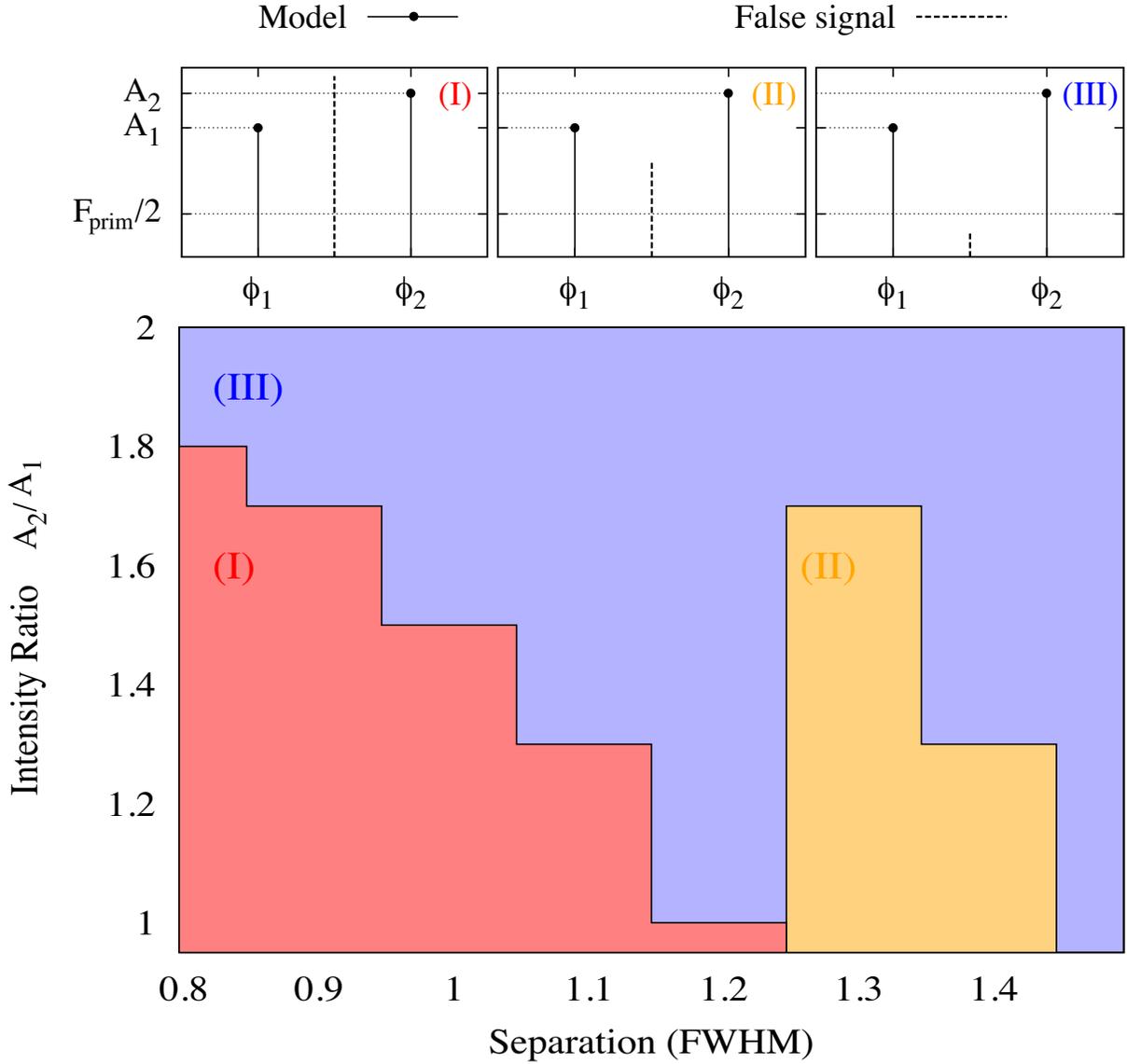}
\caption{Appearance of RM ambiguity in lower panel. (I) False signals
  appear for some intrinsic polarization angles and they are larger
  than the correct signals with intensities $A_1$ and $A_2$.  (II)
  False signals appear for some intrinsic polarization angles and they
  are smaller than the correct signals. (III) There is no false
  signals for any intrinsic polarization angle.  In upper panels, we
  show the examples of type of False signals. False signals are
  defined as the CLEAN components arisen in Faraday depth from
  $\Delta\phi/4$ to $3\Delta\phi/4$ with an amplitude larger than half
  of that of the largest CLEAN component $F_{\rm prim}$. \label{f9} }
\end{figure}

\placefigure{f9}

Figure \ref{f9} shows the distribution of the three types. There is a
tendency that larger $A_2/A_1$ reduces the generation of false signals
and two sources can be successfully resolved even for a separation
smaller than the FWHM of the RMSF if $A_2/A_1 \gtrsim 1.8$. It is seen
that false signals are serious when the two sources have comparable
intensities and the separation is about the FWHM.

%% {\bf (Note that we
%%   call all unexpected signals "false signal", although unresolved two
%%   sources which have same intrinsic polarization angle generally can
%%   be interpreted as a "single" source in terms of practical
%%   observation, that means it should not be called as "false signal".)}

\subsection{Frequency Coverage}

\begin{figure}[tp]
\figurenum{10}
\epsscale{1.0}
\includegraphics[width=160mm]{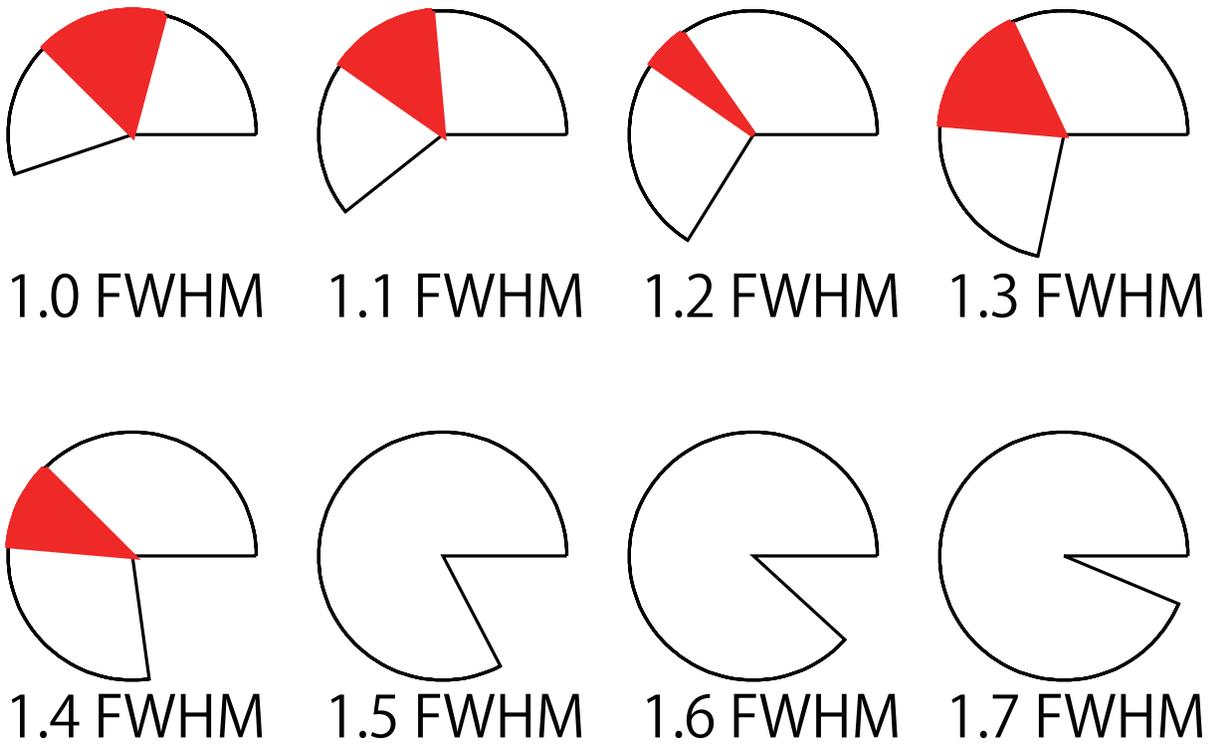}
\caption{
Same as Figure \ref{f7} but for the case of SKA. \label{f10}
}
\end{figure}

\placefigure{f10}

Finally, we change frequency coverage. Figure \ref{f10} shows the
results of the same analysis in Figure \ref{f7} but for the proposed
bandwidth of SKA. We find that the results are very similar to the
case of ASKAP. Thus, false signals are unavoidable regardless of the
bandwidth, if we scaled $\phi$ separation by the
FWHM of RMSF. We expect that false signals appear in the cases with a
certain $\Delta \chi_0$ whenever $\Delta\phi$ is less than $\sim 1.5$
FWHM.

\section{Summary and discussion}

Faraday tomography combined with RM CLEAN is a powerful method to
investigate the distribution of polarized sources along the
LOS. However, in the presence of multiple sources, large false signals
can appear in the middle of the correct signals in $\phi$ space. In
this paper, we studied the condition of the appearance of false
signals, focusing on the difference in the intrinsic polarization
angles, the separation in $\phi$ space and the intensity ratio. We
simulated the observations of ASKAP and SKA assuming two polarized
sources and found that false signals can appear for a separation as
large as 1.5 FWHM of the RMSF and the false signals exceed the correct
signals for a separation smaller than 1.1 FWHM. The intensity ratio is
also a important factor and large ratio tends to reduce the RM
ambiguity. It should be noted that these results does not depend on
the bandwidth of the telescope so that even future wide-band
observation would suffer from RM ambiguity.

It should be noted that an absolute value of $\Delta\phi$ for which we
see false signals is decreased as decreasing the FWHM. More
specifically, false signals arise for the relative RM smaller than
22.26 rad/m$^2$ with ASKAP, but than 0.189 rad/m$^2$ with
SKA. Therefore, false signals would become a minor issue in the SKA
era while we study RMs of a few rad/m$^2$ or larger.

RM CLEAN has three parameters, the gain factor $\gamma$, threshold
intensity, and maximum iteration. We also have studied the cases with
different parameters, and found that larger gain factor tends to
slightly increase the range of $\Delta\chi_0$ where false signals
arise. On the other hand, threshold intensity and maximum iteration in
this work are rather conservative values and the results were not
changed significantly. Note that threshold intensity is usually
referred to RMS noise level of the observations. Therefore, if we
study polarization sources with less than mJy, the threshold intensity
in this work could reach the RMS noise levels of ASKAP and SKA, and
thus should increase the threshold. For such cases, we may need more
careful arguments for the RM CLEAN parameters.

Finally, Stokes' QU-fitting is another powerful method to probe cosmic
magnetism with polarization observation \citep{osu12,ide13}. In this
method, fitting of polarization spectrum is performed assuming a model
of source distribution with free parameters. Combination of QU-fitting
and Faraday tomography would be effective to identify and avoid RM
ambiguity, and detect correct polarized sources. This possibility will
be pursued elsewhere in near future.

\acknowledgments

This work is supported in part by the Grant-in-Aid from the Ministry
of Education, Culture, Sports, Science and Technology (MEXT) of Japan,
No. 23740179, No. 24111710 and No. 24340048 (KT). K.K. and
T.A. acknowledges the supports of the Japan Society for the Promotion
of Science (JSPS).

\clearpage

\begin{deluxetable}{ccccc}
\tablenum{1}
\tabletypesize{\small}
\tablewidth{0pt}
\tablecaption{Frequency Coverage and the FWHM. \label{tab1}}
\tablehead{
Project & Frequency & $\lambda^2_{\rm max}$ & $\lambda^2_{\rm min}$ & FWHM \\
& GHz & m$^2$ & m$^2$ & rad/m$^2$
}
\startdata
ASKAP & $0.7-1.8$ & 1.83 $\times 10^{-1}$ & 2.76 $\times 10^{-2}$ & 22.26 \\
SKA & $0.1-10$ & 1.83 $\times 10^1$ & 9.0 $\times 10^{-4}$ & 0.189
\enddata
\end{deluxetable}

\clearpage

\clearpage

\clearpage

\clearpage

\clearpage

\clearpage

\clearpage

\clearpage

\clearpage

\clearpage

\end{document}